\documentclass{aastex}
\usepackage{spr-astr-addons}
\usepackage{url}

\RequirePackage{color}

\begin{document}

\title{Standard candles from the Gaia perspective}
\slugcomment{Not to appear in Nonlearned J., 45.}
\shorttitle{Standard candles from the Gaia perspective}
\shortauthors{Laurent Eyer et al.}

\author{Laurent Eyer}
\affil{Geneva Observatory, University of Geneva, 1290 Sauverny, Switzerland}
\and \author{Lovro Palaversa}
\affil{Geneva Observatory, University of Geneva, 1290 Sauverny, Switzerland}
\and \author{Nami Mowlavi}
\affil{Geneva Observatory, University of Geneva, 1290 Sauverny, Switzerland}
\altaffiltext{}{ISDC, Geneva Observatory, University of Geneva, 1290 Versoix, Switzerland}
\and \author{Pierre Dubath}
\affil{ISDC, Geneva Observatory, University of Geneva, 1290 Versoix, Switzerland}
\and
\author{Richard I. Anderson}
\affil{Geneva Observatory, University of Geneva, 1290 Sauverny, Switzerland}
\and
\author{Dafydd W. Evans}
\affil{Institute of Astronomy, Cambridge University, Madingley Road, Cambridge CB3 0HA, UK}
\and
\author{Thomas Lebzelter}
\affil{Department of Astronomy University of Vienna, Tuerkenschanzstrasse 17, 1180 Vienna, Austria}
\and
\author{Vincenzo Ripepi}
\affil{INAF - Astronomical Observatory of Capodimonte, Salita Moiariello 16, 80131 Napoli, Italy}
\and
\author{Laszlo Szabados}
\affil{Konkoly Observatory of the Hungarian Academy of Sciences, 1121 Budapest, Hungary}
\and
\author{Silvio Leccia}
\affil{INAF - Astronomical Observatory of Capodimonte, Salita Moiariello 16, 80131 Napoli, Italy}
\and
\author{Gisella Clementini}
\affil{INAF-Osservatorio Astronomico di Bologna, via Ranzani 1, 40127, Bologna, Italy}

\begin{abstract}

The ESA Gaia mission will bring a new era to the domain of standard candles. Progresses in this domain will be achieved thanks to unprecedented astrometric precision, whole-sky coverage and the combination of photometric, spectrophotometric and spectroscopic measurements. The fundamental outcome of the mission will be the Gaia catalogue produced by the Gaia Data Analysis and Processing Consortium (DPAC), which will contain a variable source classification and specific properties for stars of specific variability types. We review what will be produced for Cepheids, RR Lyrae, Long Period Variable stars and eclipsing binaries. 
\end{abstract}


\section{Introduction}

The subject of standard candles is a fundamental scientific case that will greatly benefit from the Gaia mission. In this respect Gaia is unique and this scientific subject will take advantage from all aspects of the mission. The astrometry will obviously provide a major contribution. However, other aspects of the Gaia measurements will contribute to this subject as well. The classical standard candles are RR Lyrae and Cepheid stars. But Gaia will also offer the possibility to exploit other classes of variable stars as standard candles. Examples of ``non-classical'' standard candles include Long Period Variables (LPVs) (\citealt{Feastetal1989}; \citealt{Matsunagaetal2009}), OGLE Small Amplitude Red Giants (OSARGs; \citealt{Wrayetal2004}), eclipsing binaries \citep{Paczynski1997} and Large Amplitude $\delta$\,Scuti stars \citep{McNamara1997}.

Certain types of stellar variability occur within specific mass and
metallicity ranges and at given evolutionary stages. The size of the Gaia
dataset will ensure that all these cases will be covered statistically.
Gaia data will allow us to establish the fundamental astrophysical
properties of these stars, in particular their luminosity, and thereby
establish their usefulness as standard candles. They will also determine their
cosmic scatter. Gaia will therefore be able to test the universality of many
different standard candles.

The astrometric precision of the Hipparcos satellite has been
exploited to its limit in the case of standard candles. Standard candles like
Cepheids or RR Lyrae can be considered ``far'' for Hipparcos precision.
The statistical properties of distance, parallax and absolute magnitude are complex.
\cite{FeastCatchpole1997} constrained the zero point of the period-luminosity relation
with a subtle and sound method. However, their zero point has been subject to
discussions. It is worth mentioning that the number of Cepheids retained for constraining their solution was only 26.
The distance modulus of the Large Magellanic Cloud (LMC), obtained with the Hipparcos-derived zero-point
is 18.7$\pm 0.10$, which seems a bit too far compared to other standard estimates that are close to 18.5 (cf. \citealt{Clementinietal2003}).

Decades after the historical discovery of the Cepheid period-luminosity
relation by Henrietta Leavitt, surveys towards the Magellanic Clouds
such as OGLE, MACHO and EROS, provided a remarkable
contribution to the subject of standard candles.
Even if the distances to the Magellanic Clouds are not known precisely, they can be
considered at a fixed distance with only moderate depth and therefore the difference
in apparent magnitude can be interpreted as a difference in luminosity. This is
why the Magellanic Clouds have offered such a high scientific potential.
Gaia, as a whole-sky survey, will observe the Magellanic
clouds, as well as the halo and the plane of the Galaxy (with parallaxes). The
inter-comparison of different populations of standard candle classes
with a single instrument will have an enourmous impact, since, for the first time, they can be calibrated to a homogeneous reference.

\section{A quick review of the Gaia mission}

In this section, we present properties of the Gaia mission that are
relevant to standard candles.

Gaia is a spacecraft of the European Space Agency (ESA) that will be
located at the Lagrangian L2 point, 1.5 million km away from Earth. It
will observe about 1 billion objects with a magnitude between
$V\simeq$~6 and 20 mag. The measurements gather astrometric, photometric,
spectrophotometric and spectroscopic data. The length of the mission
is 5 years with a possible one year extension. For a duration of 5 years,
the average number of measurements will be about 70 per object (this
number contains estimated dead-times). The launch is foreseen for
2013. There will be an alert system and intermediate data releases
throughout the mission. The final results will be made available by
2020--2021.

A summary of performances can be found on the Gaia webpage
(\url{http://www.rssd.esa.int/Gaia}) under Science Performances. The
numbers displayed in tables~\ref{tab:astroerror} and~\ref{tab:RVSperf} are extracted from this webpage (as of
October 2011).

\paragraph{The scanning law}

The Gaia scanning law has been designed in order to optimize the
astrometric results of the mission. The Gaia satellite has two fields
of view (FOV) of 0.7$\times$0.7 deg$^2$ each. These two viewing
directions are separated by an angle of 106.5 deg. The two FOV images
are superposed on the same focal plane that consists of
106 CCDs, totaling nearly 1 billion pixels. As the satellite rotates
around its axis with a period of 6 hours, the stars are sweeping through the
focal plane. The CCDs are read in Time Delay Integration
mode. The rotation axis of the satellite is precessing on a Sun-centered
cone with an opening angle of 45 degrees and a precession
period of 63 days. This constant angle of the rotation
axis with respect to the Sun's position gives the peculiar dependency of
the scanning law on ecliptic coordinates. The average number of
transits (one passage through the FOV) is about 70 but varies between 40
and 250, depending on the sky position.

Properties of the scanning law have been presented by \cite{EyerMignard2005}. Although there have been changes in satellite design, the conclusions of the study remain valid for the Astrometric Field (AF). The spectral window of Gaia sampling contains high peaks at high frequencies, limiting aliasing when compared to large scale ground-based surveys (see \citealt{Eyeretal2009}). The period recovery of periodic signals is very high even at relatively low signal to noise ratio (see \citealt{EyerMignard2005}). Handling of semi-regular or irregular variables might be difficult due to the sparsely sampled Gaia time series. Furthermore, the semi-regular or irregular variables may contaminate the samples of periodic objects. Particular  cases such as double-mode Cepheids or Blazhko RR Lyrae stars will also require special analysis.

\paragraph{The astrometric performance}

The mission's success depends critically on the astrometric performance achieved.
Therefore, performance estimations
have been the subject of constant attention throughout the
development of the spacecraft. The latest numbers available, re-evaluated
during the critical design review in April 2011, are displayed in
Table~\ref{tab:astroerror}. 

These numbers result from studies which have been recently quite stable; the current numbers are consistent with those obtained in the past few years. Globally, the performance is within the initial requirements with only some minor non-compliances.

The numbers shown in Table~\ref{tab:astroerror} represent estimated errors on
parallax at the end of the mission. They should be multiplied by 0.8 and 0.5
in order to obtain errors for end-of-mission position proper motion ($\mu$as/year), respectively.

\begin{table}[t]
\caption{Astrometric error for end-of-mission parallax as a function of spectral type and magnitude}
\label{tab:astroerror}
\begin{tabular}{@{}lccc}
\hline
& B1V &  G2V & M6V  \\ \hline \hline
$V-I$ & $-$0.22 &  0.75 & 3.85 \\ \hline
Bright stars & \tiny{$6<V<12$} &  \tiny{$6<V<12$} &  \tiny{$8<V<14$} \\
& 5-14$\,\mu$as &  5-14$\,\mu$as & 5-14$\,\mu$as \\ \hline
V=15 &  26$\,\mu$as &  24$\,\mu$as & 9$\,\mu$as    \\ \hline
V=20 & 330$\,\mu$as &  290$\mu$as & 100$\,\mu$as  \\
\hline
\end{tabular}
\end{table}

In theory, the performance improves with mission length $L$. The
parallax and position errors scale as $L^{-0.5}$ while the proper motion
error varies as $L^{-1.5}$. However, the astrometric
solution of the first 18 months is likely to be affected by systematic errors. The goal of the
consortium will be to search and correct for these errors through the
subsequent iterations in such a way as to reach the expected accuracy at
the end of the mission. As a consequence, the scaling of the different
solutions produced during the mission may not strictly follow the above
rules.

For a Cepheid located at 12 kpc with a 10 day period, a relative parallax 
error of 10\% is expected assuming no extinction. The same relative error is
reached for a 6 kpc Cepheid if the extinction is $A_V=5$ mag. For Cepheids, Gaia
will cover a significant fraction of the Galaxy with a very good precision.

\paragraph{The photometric and spectrophotometric performance}

\begin{figure*}[t]
\includegraphics[angle=90,width=16cm]{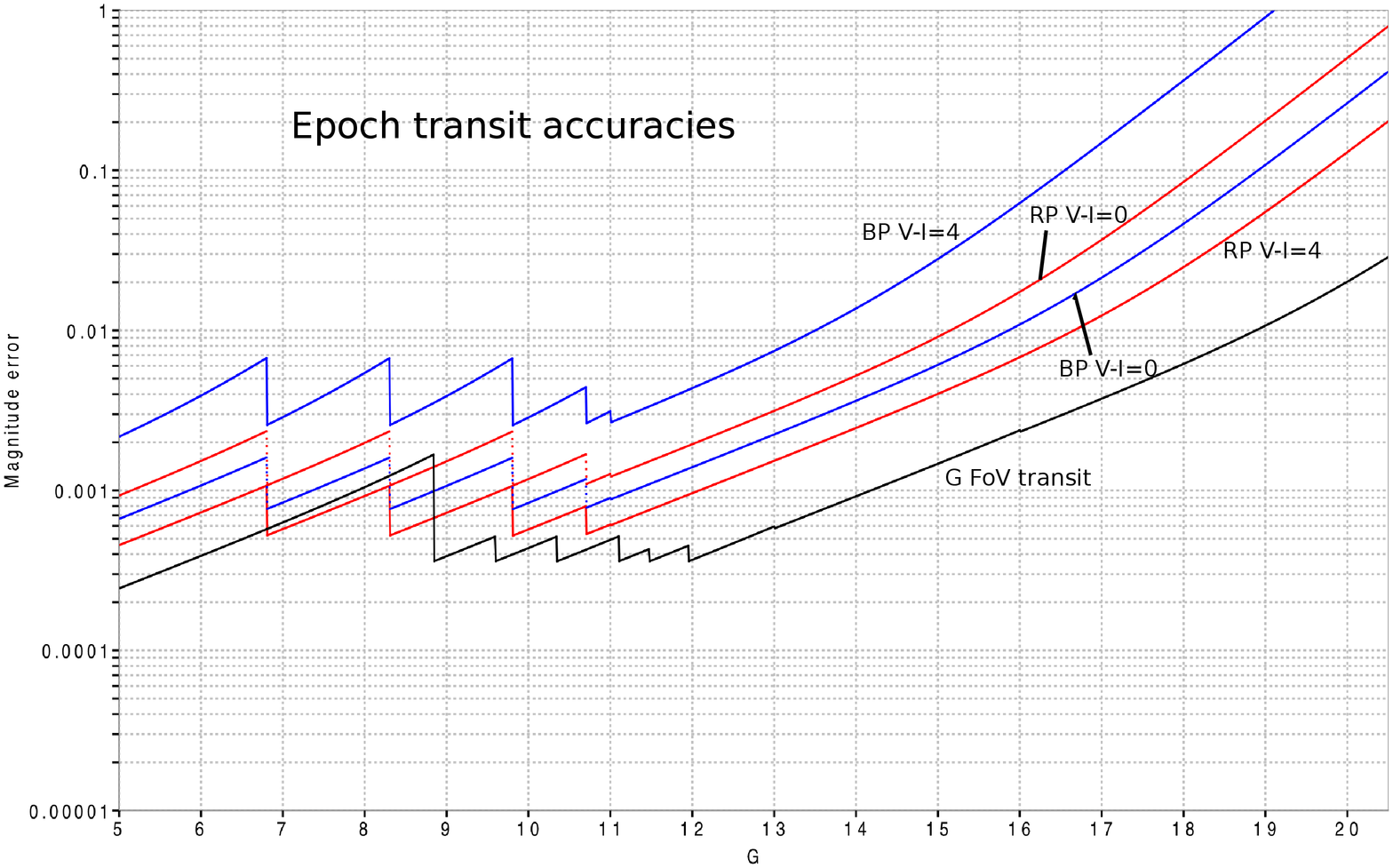}
\caption{\label{fig:epochphotom} %
 Per transit photometric error in G band, integrated BP and RP as a
 function of G magnitude. The sawtooth structure at the bright
 magnitudes is a consequence of the gating system which allows
 observations of bright stars by limiting the exposed part of the
 CCD, thus reducing the integration time} 
\end{figure*}

The astrometric field is also producing a white light (called G-band)
magnitude. As there is no filter, the bandpass is only limited by the
optical properties of the system (reflectivity of the mirror, response
of the CCDs, etc.). The wavelength coverage is from 330 to
1050\,nm. The photometric G-band precision should be of very high
quality, as it is the sum of the 9 CCD measurements over one FOV
transit. The transit/epoch photometry accuracy as a function of the
magnitude is given in Figure~\ref{fig:epochphotom}.
The lower limit of the calibration error is estimated to be at the level of 1 mmag. The per-CCD
photometry will also be available and will allow detection of variability on
very short time scales (tens of seconds).

Gaia performs also low resolution spectrophotometry. The Blue Photometer
(BP) covers the wavelength range from 330 to 680\,nm while the Red
Photometer (RP) covers the range 640--1050\,nm. The RP has
red-enhanced CCDs so that longer wavelengths are reached. The
requirements were not formulated in terms of acquiring desired astrophysical
quantities, but in terms of photometric precision in
pseudo-bands. The BP and RP
spectra have each 60 samples. The photometric precision of the
integrated or ``mean" spectra are given in
Figure~\ref{fig:epochphotom}.  However, again, these numbers should be
seen as theoretical limits. The lower limit of the calibration error is estimated to be at the
level of 10 to 30\,mmag. It should be noted that the error estimations
from the April 2011 review have larger errors for this lower limit.
Finally, the Sky-Mapper (SM) CCDs will also produce photometric
measurements.  The mean number of transits in G, BP and RP is
estimated to reach 70 over 5 years.

\paragraph{The Radial Velocity Spectrometer performance}

The Radial Velocity Spectrometer (RVS) is a near-infrared instrument
with a resolution of 11,500 and spans the wavelength range from 847
to 874 nm, which covers the Calcium triplet. The RVS instrument will
survey the whole-sky up to magnitude $V \sim$ 15 or 17 (depending on the
spectral type of the stars) with end-of-mission error levels from 1 to 10 km\,s$^{-1}$, depending on the
spectral type and magnitude, cf. Table~\ref{tab:RVSperf}. The number
of measurements in the RVS will be about 40 per object over the 5 year
mission. The number of transits for the RVS is reduced, since there are only 4 dedicated CCDs perpendicular to the scanning direction
whereas the SM, G, BP, RP instruments have 7.

\begin{table}[t]
\caption{Radial velocity end-of-mission error as a function of
spectral type and magnitude}
\label{tab:RVSperf}
\begin{tabular}{@{}lrr}
\hline
Spectral type & V &  Radial velocity error  \\ \hline \hline
B1V &    7 &   1\,km\,s$^{-1}$   \\ &   12 &   9\,km\,s$^{-1}$   \\ \hline
G2V &   13 &   1\,km\,s$^{-1}$   \\ &   16.5 &  13\,km\,s$^{-1}$   \\ \hline
K1III-MP &   13.5 &   1\,km\,s$^{-1}$   \\ (metal-poor) &   17 &  13\,km\,s$^{-1}$   \\
\hline
\end{tabular}
\end{table}

\section{Standard Candles}

The knowledge about standard candles will benefit from all aspects of the Gaia mission: its astrometry, its photometry and spectrophotometry, as well as its spectrometric radial velocity measurements.

The astrometry will allow calibration of luminosities of the standard candles thanks to the Gaia parallaxes. For a given variability type, there exists an interplay between luminosity, distance (distribution within the Galaxy) and Gaia precision for the corresponding apparent magnitudes. However the number of standard candles of a given type with good and useful astrometry will increase by one to several orders of magnitude with respect to the present situation. 

The Gaia multi-epoch photometry is also advantageous in the case of standard candles with large amplitudes. For these cases, light curves can be modeled and mean luminosities can be defined.
Finally, with uniform and homogeneous photometric and spectrophotometric measurements of Gaia, calibrations of period-luminosity-color relations can be established.

Another less often mentioned benefit from  astrometry is that Gaia will be able to determine the orbit of astrometric binary stars from their movement on the sky. With radial velocity measurements also from Gaia, the physical orbit can be determined (e.g. \citealt{Zwahlenetal2004}). This is another way to determine distances purely geometrically.

Furthermore Gaia will conduct a global survey, collecting photometry and spectrophotometry on a multi-epoch basis that will allow the detection of new objects in each standard candle category, e.g. new Cepheids, new RR Lyrae stars and eclipsing binary systems, see Table~\ref{tab:prediction4Gaia}.
Due to its diversity the impact of such a harvest is difficult to forecast. However, the physics driving the variability and instability, and in particular
how metallicity affects the variability properties of a star will be systematically studied. Both aspects are essential for the calibration of standard candles using Gaia astrometry. In addition, we may find entirely new classes of standard candles within the Gaia data. 

Spectrophotometry will also provide estimates of stellar parameters for the standard candles.

Radial velocity data that will be obtained for the most luminous objects will allow computation of physical parameters of single star pulsators, using the Baade-Wesselink method and orbital parameters of binary systems with the  Wilson-Devinney-like code (see Section~\ref{SOSEB}).

The detection of new standard candle objects and characterization of their astrophysical parameters will provide a wealth of data to test, on a statistical level, the universality of standard candle relations such as the period-luminosity-metallicity relation.
Once calibrated to high accuracy with Gaia, standard candles can be used to extend studies of Galactic and extragalactic structure beyond the astrometric performance capabilities of Gaia. As examples, we mention tracing of the galactic bar using OSARGs (\citealt{Wrayetal2004}), or the possibility to constrain the three-dimensional structure of the Large and Small Magellanic Clouds by using period-luminosity relations of pulsating red giants (\citealt{Lahetal2005}). Synergies of Gaia with other large surveys such as LSST are also easily envisioned. The RR Lyrae stars calibrated with Gaia can be used by LSST to fully characterize the halo of our Galaxy.

Different standard candles correspond to stars at different stages of evolution. If the evolutionary stages of standard candles are known, the formation history of the Galaxy can be traced (see e.g. \citealt{Clementini2011}).

Another interesting application of standard candles is in mapping the distribution of the interstellar medium.

Despite many studies devoted to variable stars on one side and to populations of stars on the other side, not many studies have so far been devoted to the combination of both fields, i.e. the study of variable stars in relation to stellar populations. Here are two examples of questions showing the need for more studies in this domain.
A basic question that arises when talking about variable stars in stellar populations concerns the fraction of stars that are expected to vary. The answer is not obvious, as attested by the difficulty to predict the number of variable objects expected to be observed in a specific survey, see Table~\ref{tab:prediction4Gaia}. For example, the number of eclipsing binaries predicted to be detected by Gaia varies from 0.5 million (\citealt{DischlerSoederhjelm2005}) to 7 million (\citealt{Zwitter2002}), which represents a factor of ten uncertainty.
Moreover, it is not always possible to predict whether a star at a given location in the HR-diagram will pulsate or not. Stars are expected to pulsate in some specific areas of the HR-diagram, called instability strips, e.g. the ``classical instability strip'' is hosting Cepheids, RR Lyrae or $\delta$\,Scuti stars. The identification of the borders of those instability strips has been a successful tool to better understand the pulsational mechanism and the parameters driving the photometric variability at the surface of the stars. The crucial role of convection and of its coupling with pulsation has, for example, been highlighted in explaining the red border of the Cepheid instability strip. It explains why stars located in the HR-diagram between the instability strip and the red giant branch do not pulsate. However, what remains unexplained is why some stars observed in the instability strip do not show the expected photometric variability.
The comparison of certain types of variable stars in different known stellar populations, such as our Galaxy, open or globular clusters, or in the local group of galaxies, provides a useful way to learn about those variable stars as a function of different properties such as the metallicity. Open clusters, for example, are natural laboratories for this, as their member stars are assumed to share the same age, initial chemical abundances, distance and reddening. In this way, the already listed 2100 open clusters in the disk of our Galaxy (\citealt{Diasetal2002}), spanning a large interval in age and Galacto-centric distances, have been used as an excellent tool to probe – both the chemical and dynamical – structure and the evolution  of the Galactic disk. The variety of variable star content from one cluster to another, associated to the specific variability properties characterizing each phase of evolution, provides independent measurements for the physical parameters of open clusters. Yet, this connection  is not well known so far. One difficulty is the determination of the membership of the cluster. Gaia will put the question of membership on solid ground (see \citealt{vanLeeuwen2009} for what has been done with Hipparcos).

In Table~\ref{tab:prediction4Gaia} we review the number of discovered objects by different large scale surveys.
The estimates by \cite{EyerCuypers2000} are only for the Galaxy or some component of the Galaxy. The large uncertainty in these numbers shows that there are many unknowns in this domain.
From Table~\ref{tab:prediction4Gaia}, the optimist's view would be that Gaia will multiply by nearly 5 the number of Galactic RR Lyrae stars, by 10 the number of Galactic Cepheids, by nearly 40 the Galactic LPVs and by nearly 1000 the Galactic eclipsing binaries. To this table variables from the LMC and SMC are added.
Due to the bright limit of Gaia at V$\sim$6 mag, only few Cepheids will be missed (in the Hipparcos catalogue we found 22 stars for which the maximum light is brighter than 6).

\begin{table*}
\small
\caption{\label{tab:prediction4Gaia} Numbers of RR Lyrae stars, Cepheids, Long Period Variable stars, eclipsing binaries, known in the Galaxy, the LMC and SMC and predicted numbers in the Galaxy for the Gaia mission. For Hipparcos the numbers are taken from \cite{ESA1997}; the numbers from ASAS should be take with care (see \citealt{Berdnikovetal2009}, variable types have been selected with the ``only'' option) and are taken from the ASAS webpage as of May 2011, the estimates for Gaia are from \cite{EyerCuypers2000}; The ``other'' line are other publications:  Galactic Cepheids (\citealt{Fernieetal1995}), Gaia Cepheids (\citealt{Windmarketal2011}), Gaia eclipsing binaries (\citealt{DischlerSoederhjelm2005}, \citealt{Zwitter2002}), OGLE-III SMC RR Lyrae stars (\citealt{Soszynskietal2010b}), OGLE-III LMC RR Lyrae stars (\citealt{Soszynskietal2009a}), OGLE-III bulge RR Lyrae stars (\citealt{Soszynskietal2011}), OGLE-III SMC Cepheids (\citealt{Soszynskietal2010a}), OGLE-III LMC Cepheids (\citealt{Soszynskietal2008}), EROS LMC LPV (\citealt{Spanoetal2011}), OGLE-II SMC eclipsing binaries (\citealt{Wyrzykowskietal2004}), OGLE-III LMC eclipsing binaries (\citealt{Graczyketal2011}) }
\begin{tabular}{@{}lrrrrr@{}}
               &                 & RR Lyrae              & Cepheid     & LPV     & Eclipsing bin.\\ \hline
{\bf Known}    &                 &                       &             &         &               \\
               & Hipparcos       & 186                   &         273 &   1,238 & 917           \\
               & ASAS            & 1,635                 &         872 &   2,793 & 5,911         \\
               & Other           & (bulge) 16,839        &         509 &         &               \\ \hline
{\bf Predicted}&                 &                       &             &         &               \\ 
               & Eyer \& Cuypers & (bulge) 15,000-40,000 & 2,000-8,000 & 200'000 & 3,000,000     \\
               &                 & (halo) 70,000         &             &         &             \\
               & Other           &                       &       9,000 &         & 500,000; 7,000,000 \\ \hline
LMC            &                 & 24,906                &       3,361 &  37,047 & 26,121        \\
SMC            &                 &  2,475                &       4,630 &         &  1,351        \\
\tableline
\end{tabular}
\end{table*}

A word on Supernovae: Gaia will have an alert system that will detect supernovae and other transient events. The estimated number of supernovae brighter than magnitude 19 is around 6,000. One third of these will be observed before their maximum. This subject is covered by Gilmore in this volume.

\section{Gaia Data Processing and Analysis Consortium activities on standard candles}

Software development is often underestimated in large scale projects.
For Gaia, however, it was recognized early-on that the software development is a key element for
its success. Indeed, the Gaia data processing and analysis is a
tremendous task, due to the large amount of raw data to be
processed (in the order of 100 compressed Terabytes in 5 years), but
even more so due to the complex and interwined relationships between astrometry, photometry,
spectrophotometry and spectroscopy.
In addition, since the targeted accuracy is higher than
anything ever obtained before for so many stars, the processes have to be
self-calibrating, going through a number of iterations, with each set of
results providing inputs for the next run. This is most obvious for
the astrometric global iterative solution. The global reference frame
of sky positions will be built gradually, measuring/modeling
parallaxes and proper motions, and eliminating
deviating points such as multiple stars.

The task of the variability analysis and processing was given to the Coordination Unit~7 (CU7),
where the work is decomposed into several steps.

\paragraph{Step 1: Variability detection}
The photometric (CU5) and spectroscopic (CU6) groups are in charge of detecting
variable objects applying general-purpose algorithms, such as some
statistical standard tests. The Special Variability Detection (within
CU7) is defined to implement specific algorithms which take advantage
of what we know about a particular type of variability. All variable
objects are then stored into the Variability Database.

\paragraph{Step 2: Variability characterization}
Once variable object candidates are identified, their behavior is characterized. A number of {\em attributes} are computed to characterize the sources.  Some of them reflect global stellar properties, such as mean color or absolute magnitude, whereas others describe some of the light curve features. A number of statistical parameters are derived from the magnitude distribution.
Since we know that Gaia has a relatively good performance on periodic objects, a period search is carried out and the folded light curves are modeled with Fourier series. Many harmonics are fitted, but only those that are significant according to an F-test are kept. The number of harmonics is also limited if there are gaps in the time sampling to avoid non-physical large model excursions in regions devoid of measurements.

\paragraph{Step 3: Variability classification}
Once variables are identified and characterized, multiple classification methods are applied.
We decompose this into three
subtasks: supervised methods, clustering techniques, extractors (a
specific variability type is selected using all the astrophysical
knowledge).  Automated and efficient variable star detection and
classification are critical components of large-scale surveys. They
are required both to study stellar population properties and to
provide candidates for further detailed investigation of individual
cases. Tests have been performed for supervised methods by using
cleaned Hipparcos light curves to evaluate the ultimate performance of
periodic (\citealt{Dubathetal2011}) and non-periodic (Rimoldini et al., in
preparation) star classification schemes. The classification result 
associates a given variable star with a membership probability to 
a given class of variable stars.

\paragraph{Step 4: Specific Object Studies}
In the three first tasks, data for all objects are processed in a
systematic way. In Specific Object Studies, specific algorithms are
applied to objects as a function of their variability class.
For example, the processing required at this stage for
the periodic Cepheid stars is different from the one required for the
usually rather erratic distant Active Galactic Nuclei. After the Specific
Object Studies step, all available information about the variables has been
extracted from the Gaia data and is available in the Variability
Database.

We will be able to validate the automated classification, in this step, by analyzing the objects in greater details and studying the properties of sub-samples (sometimes manually). The Specific Object Studies step should also re-evaluate the membership probability assigned by the automated classification.

We  present what is foreseen for the Specific Object Studies of standard candles, namely in section~\ref{SOSRRCEP} for RR Lyrae stars and Cepheids, in section~\ref{SOSLPV} for Long Period Variables, in section~\ref{SOSEB} for Eclipsing Binaries and in section~\ref{SOSSN} for Supernovae.

\paragraph{Step 5: Global Variability Studies}
In the next step, the Global Variability Studies task will investigate larger-scale properties of variability, e.g. the period distribution for all Cepheids, a color-magnitude (or HR) diagram with iso-contours of variability amplitude, etc. Given the large number of objects, special tools are in development in order to facilitate the evaluation and usage of the database content.

\subsection{Cepheids and RR Lyrae stars}
\label{SOSRRCEP}
A good characterization of Cepheids and RR Lyrae stars is essential if
those objects are to be used as standard candles.  The automatic data
processing implemented in the CU7 pipeline should first derive the
period(s) and the Fourier decomposition parameters of the light
curves, i.e.\:the amplitude ratios and phase differences (see \citealt{Lecciaetal2011}).  These parameters allow the identification
of the pulsation mode(s), fundamental or higher overtones.  They will
also be useful to {\bf estimate} the metal abundance of those objects and
their intrinsic colors.

Double-mode RR Lyrae variables will be detected and the period
ratio fundamental/first overtone used to estimate the mass of the
targets. \par
A significant sample of RR Lyrae (about 30\%) are expected to be
affected by the Blazhko effect, which is a modulation of the
amplitude and/or phase of the periodic signal, that can lead to an
erroneous determination of mean magnitude and stellar parameters.
The detection of Blazhko RR Lyrae and the estimate of the Blazhko
period (when possible) has been implemented in the CU7 processing chain.

For Cepheids, the pipeline will be able to distinguish between
population I (Classical Cepheids) and population II objects
(BL Her, W Vir classes) that obey different Period-Luminosity or Period-Luminosity-Color relations.\par
In this context, the question of binarity is also important.  The 
presence of
a companion leads to smaller photometric amplitudes and an altered mean luminosity
compared to the light curves of single Cepheids
(e.g. \citealt{KlagyivikSzabados2009}).  This would lead to an
erroneous distance estimate from the period-luminosity relation,
if the effect of the companion is
not taken into account.

Stellar parameters can be derived with the knowledge of multi-epoch
radial velocities, measured by Gaia. The radii and the distances of 
Cepheids and RR
Lyrae can be derived with the Baade-Wesselink (BW) method, with an error of
a few \% for bright objects and about 10\% at V magnitudes of 13-14.
The resulting  quantities can be compared with the values obtained
through Gaia astrometry, providing stringent constraints on the
systematic errors affecting the method. In particular, it will be
possible to fix the value of the ``projection factor p'', a
proportionality constant between
pulsational and radial velocities, which is the most important source of 
error
for any version of the BW technique. This will allow to safely use the BW method for faint
objects outside the Galaxy.
\subsection{Long Period Variables}
\label{SOSLPV}
Long Period Variables (LPVs) constitute a class of red giant variables classically defined by the Mira stars and semi-regular variables.
These variables are known to obey several nearly parallel period-luminosity relations related to the various pulsation modes.
Miras are known to be fundamental pulsators, and variables with smaller amplitudes are typically first or second overtone pulsators.
Their P-L relations have been identified from observations, thanks to large scale surveys initiated by MACHO (\citealt{Woodetal1999}), OGLE (\citealt{Soszynskietal2009b}), and EROS (see e.g., \citealt{Spanoetal2011}).
Unfortunately, the modeling of the pulsation of those red giants is difficult due to the coupling between pulsation and convection.
The agreement between predicted and observed relations is quite good, but the impact of stellar parameters on the pulsation properties is
still poorly understood. Large scale surveys with multi-epoch photometry of red giants in the Magellanic Clouds, the distances to which are known, 
have been the driving tool to study the P-L relations of LPVs.

By providing the distances to a large set of red giants in our Galaxy, Gaia will bring substantial additional data for those studies,
widening the range of stellar parameters and relating the zero points of the relations to a direct measurement. It will also be possible to
study various subgroups within the Milky Way and to place nearby and well-studied LPVs onto a distinct P-L relation.
This study, however, requires the knowledge of the bolometric correction for each star, the value of which strongly depends on the atmospheric
chemistry (O or C-rich, with or without dust). The O or C-rich nature of the star can be assessed from the relative strengths of the TiO and CN 
molecular bands around 7780 and 8120~\AA, which fall within the spectral range of the BP spectrometer of Gaia (6500-10000~\AA).
First investigations to see whether the resolution and sensitivity of the spectrometer is sufficient to allow  for a good distinction into O or C-rich stars
led to promising results awaiting fine tuning once Gaia is in orbit.

Furthermore, special care has to be taken on a possible shift of the photo-center of red giants due to their large radii and the possibility of large
scale and variable structures on their surface, e.g. as the result of convection. The impact on the astrometry of Betelgeuse in the Hipparcos $H_p$ band, for example, is estimated to be of the order of 3.4~mas (\citealt{Harperetal2008}). This effect in Gaia is investigated by the Coordination Unit~4 in the Gaia Consortium.


\subsection{Eclipsing binaries}
\label{SOSEB}
Eclipsing binaries are traditionally sub-classified as EA, EB or EW types.
EA, or Algol ($\beta$-Persei), types have the eclipses well defined in their light curves, with the possibility to identify the times of their beginning and their end in the folded light curve.
EB, or $\beta$-Lyrae, types display a continuous variation of the light curve over an orbital cycle, preventing the identification of the eclipse times.
EW, or W Ursae Majoris, types have similar depths of the primary and secondary eclipses.

\cite{Pojmanski2002} suggested a classification based on the physical property of the binary, i.e. detached, semi-detached or contact binary.
These three categories can theoretically be identified from the $a_2$ and $a_4$ parameters of the Fourier decomposition of the light curves in cosine series (\citealt{Rucinski1993}, \citealt{Pojmanski2002}).

The automated variability processing pipeline for Gaia will also characterize the geometry of the folded light curves of eclipsing binaries and estimate the duration of the eclipses and their depths and phases. A study is underway to explore the orbital parameters that can be estimated based solely on the geometrical characterization of the folded light curves in order to help the sub-classification. Initial guesses of some of the parameters may also be helpful for the computation of the full parameters of the binary system with a Wilson-Devinney-type code, a task in the hand of Coordination Unit~4 in the Gaia consortium.

\subsection{Supernovae}
\label{SOSSN}
The light curves of cataclysmic variables and supernovae will also be characterized by CU7 and the results will be made available in the Gaia catalogue of variable stars.
Contrary to the alert system of CU5 whose purpose is to alert the scientific community as early as possible and hence is derived from basically calibrated data, the data analyzed by CU7 will consist of fully calibrated data that cover the light curve of supernovae over the entire duration available at the time of catalogue publication.

\section{Conclusions}
Gaia's scientific impact on standard candles will be remarkable. Clearly, there will be some limitations related to statistical aspects of the analysis: aliasing in the period search and the estimations of the circumstellar or interstellar extinction as well as the bolometric correction.
However, Gaia will calibrate many standard candles and these will contribute to many different topics. In this article we touched on a few: testing the universality of standard candles, deriving statistical properties of different types of standard candles, searching for new standard candles, constraining stellar evolution through standard candles, improving the knowledge of the formation history of the Galaxy and extending Gaia's results to investigate Galactic structure. Some of the problems might be used to our advantage and some standard candles could be utilized to crosscheck Gaia's astrometry or even to establish extinction maps (\citealt{Windmarketal2011}).
 
\nocite{*}

\bibliographystyle{spr-mp-nameyear-cnd}

\begin{thebibliography}{}
\bibitem[Berdnikov et al.(2009)]{Berdnikovetal2009} Berdnikov, L.~N., Kniazev, A.~Y., Kravtsov, V.~V., Pastukhova, E.~N., \& Turner, D.~G.\ 2009, Astronomy Letters, 35, 39
\bibitem[Clementini et al.(2003)]{Clementinietal2003} Clementini, G., Gratton, R., Bragaglia, A., et al.\ 2003, \aj, 125, 1309
\bibitem[Clementini (2011)]{Clementini2011} Clementini, G. \ 2011, EAS Publications Series, 45, 267
\bibitem[Dischler \& Soederhjelm(2005)]{DischlerSoederhjelm2005} Dischler, J., Soederhjelm, S.\ 2005, The Three-Dimensional Universe with Gaia, 576, 569
\bibitem[Dias et al.(2002)]{Diasetal2002} Dias, W.~S., Alessi, B.~S., Moitinho, A., \& L{\'e}pine, J.~R.~D.\ 2002, \aap, 389, 871
\bibitem[Dubath et al.(2011)]{Dubathetal2011} Dubath, P., Rimoldini, L., S{\"u}veges, M., et al.\ 2011, \mnras, 414, 2602 
\bibitem[ESA(1997)]{ESA1997} ESA, 1997, The Hipparcos and Tycho Catalogues, ESA SP-1200 
\bibitem[Eyer \& Cuypers(2000)]{EyerCuypers2000} Eyer, L., \& Cuypers, J.\ 2000, IAU Colloq.~176: The Impact of Large-Scale Surveys on Pulsating Star Research, 203, 71 
\bibitem[Eyer \& Mignard(2005)]{EyerMignard2005} Eyer, L., \& Mignard, F.\ 2005, \mnras, 361, 1136 
\bibitem[Eyer et al.(2009)]{Eyeretal2009} Eyer, L., Mowlavi, N., Varadi, M., et al.\ 2009, SF2A-2009: Proceedings of the Annual meeting of the French Society of Astronomy and Astrophysics, 45
\bibitem[Fernie et al.(1995)]{Fernieetal1995} Fernie, J.~D., Evans, N.~R., Beattie, B., \& Seager, S.\ 1995, Information Bulletin on Variable Stars, 4148, 1
\bibitem[Feast et al.(1989)]{Feastetal1989} Feast, M.~W., Glass, I.~S., Whitelock, P.~A., \& Catchpole, R.~M.\ 1989, \mnras, 241, 375 
\bibitem[Feast \& Catchpole(1997)]{FeastCatchpole1997} Feast, M.~W., \& Catchpole, R.~M.\ 1997, \mnras, 286, L1
\bibitem[Graczyk et al.(2011)]{Graczyketal2011} Graczyk, D., Soszy{\'n}ski, I., Poleski, R., et al.\ 2011, \actaa, 61, 103 
\bibitem[Harper et al.(2008)]{Harperetal2008} Harper, G.~M., Brown,  A., \& Guinan, E.~F.\ 2008, \aj, 135, 1430
\bibitem[Klagyivik \& Szabados(2009)]{KlagyivikSzabados2009} Klagyivik, P., \& Szabados, L.\ 2009, \aap, 504, 959 
\bibitem[Lah et al.(2005)]{Lahetal2005} Lah, P., Kiss, L.~L., \& Bedding, T.~R.\ 2005, \mnras, 359, L42 
\bibitem[Leccia et al.(2011)]{Lecciaetal2011} Leccia, S., Ripepi, V., Clementini, G., et al.\ 2011, in:  The Fundamental Cosmic  Distance Scale: State of the Art and the Gaia Perspective, on line proceedings, available on: http://www.na.astro.it/ESFdistance/ 
\bibitem[McNamara(1997)]{McNamara1997} McNamara, D.\ 1997, \pasp, 109, 1221 
\bibitem[Matsunaga et al.(2009)]{Matsunagaetal2009} Matsunaga, N., Kawadu, T., Nishiyama, S., et al.\ 2009, \mnras, 399, 1709 
\bibitem[Paczy{\'n}ski(1997)]{Paczynski1997} Paczy{\'n}ski, B.\ 1997, The Extragalactic Distance Scale, edited by M. Livio (Cambridge University Press) 273
\bibitem[Pojmanski(2002)]{Pojmanski2002} Pojmanski, G.\ 2002, \actaa, 52, 397 
\bibitem[Rucinski(1993)]{Rucinski1993} Rucinski, S.~M.\ 1993, \pasp, 105, 1433 
\bibitem[Soszy{\'n}ski et al.(2008)]{Soszynskietal2008}  Soszy{\'n}ski, I.,  Poleski, R., Udalski, A., et al.\ 2008, \actaa, 58, 163
\bibitem[Soszy{\'n}ski et al.(2009a)]{Soszynskietal2009a} Soszy{\'n}ski, I., Udalski, A., Szyma{\'n}ski, M.~K., et al.\ 2009, \actaa, 59, 1 
\bibitem[Soszy{\'n}ski et al.(2009b)]{Soszynskietal2009b} Soszy{\'n}ski, I., Udalski, A., Szyma{\'n}ski, M.~K., et al.\ 2009, \actaa, 59, 239
\bibitem[Soszy{\'n}ski et al.(2010a)]{Soszynskietal2010a} Soszy{\'n}ski, I., Poleski, R., Udalski, A., et al.\ 2010, \actaa, 60, 17
\bibitem[Soszy{\'n}ski et al.(2010b)]{Soszynskietal2010b} Soszy{\'n}ski, I., Udalski, A., Szyma{\'n}ski, M.~K., et al.\ 2010, \actaa, 60, 165
\bibitem[Soszy{\'n}ski et al.(2011)]{Soszynskietal2011}  Soszy{\'n}ski,  I., Dziembowski, W.~A., Udalski, A., et al.\ 2011, \actaa, 61, 1 
\bibitem[Spano et al.(2011)]{Spanoetal2011} Spano, M., Mowlavi, N.,  Eyer, L., et al.\ 2011, arXiv:1109.6132
\bibitem[van Leeuwen(2009)]{vanLeeuwen2009} van Leeuwen, F.\ 2009, \aap, 497, 209 

\bibitem[Windmark et  al.(2011)]{Windmarketal2011} Windmark, F., Lindegren, L., \& Hobbs, D.\ 2011, \aap, 530, A76 
\bibitem[Wood et al.(1999)]{Woodetal1999} Wood, P.~R., Alcock, C., Allsman, R.~A., et al.\ 1999, Asymptotic Giant Branch Stars, 191, 151
\bibitem[Wray et al.(2004)]{Wrayetal2004} Wray, J.~J., Eyer, L., \& Paczy{\'n}ski, B.\ 2004, \mnras, 349, 1059 
\bibitem[Wyrzykowski et al.(2004)]{Wyrzykowskietal2004} Wyrzykowski, L., Udalski, A., Kubiak, M., et al.\ 2004, \actaa, 54, 1 
\bibitem[Zwahlen et al.(2004)]{Zwahlenetal2004} Zwahlen, N., North, P., Debernardi, Y., et al.\ 2004, \aap, 425, L45
\bibitem[Zwitter(2002)]{Zwitter2002} Zwitter, T.\ 2002, Exotic Stars as Challenges to Evolution, eds. C.A. Tout \& W. van Hamme, ASPC 279, 31 
\end{thebibliography}

\end{document}